\documentclass[a4paper]{jpconf}
\usepackage{graphicx}
\begin{document}
\title{Diffractive Optics for Gravitational Wave Detectors}
\author{A Bunkowski$^1$, O Burmeister$^1$, T Clausnitzer$^2$, E-B Kley$^2$, \\A T\"unnermann$^2$, K Danzmann$^1$ and R Schnabel$^1$}
\address{$^1$Max-Planck-Institut f\"ur Gravitationsphysik
 (Albert-Einstein-Institut) and\\ Universit\"at Hannover,
 Callinstr. 38, 30167 Hannover, Germany}
\address{$^2$Institut f\"ur Angewandte Physik,
Friedrich-Schiller-Universit\"at Jena, Max-Wien-Platz 1, 07743
Jena, Germany }
\ead{alexander.bunkowski@aei.mpg.de}
\begin{abstract}
All-reflective interferometry based on nano-structured diffraction
gratings offers new possibilities for gravitational wave
detection.
We investigate an all-reflective Fabry-Perot interferometer
concept in 2nd order Littrow mount.
The input-output relations for such a resonator are derived
treating the grating coupler by means of a scattering matrix
formalism.
A low loss dielectric reflection grating has been designed and
manufactured to test the properties of such a grating cavity.
\end{abstract}
\section{Introduction}
Laser interferometric gravitational wave detectors employ partly
transmissive mirrors as 50/50 beam splitters and couplers to
cavities.
To avoid thermal effects associated with laser power absorption in
transmitted mirror substrates all-reflective
interferometer-topologies can be used~\cite{Sun97}.
All-reflective interferometers have the additional advantage that
opaque materials with potentially superior mechanical properties,
e.g. silicon~\cite{Rowan}, can be used as mirror substrates.

Previously realized all-reflective cavity concepts require high
1st order diffraction efficiency for high finesse cavities.
Here we report on the investigation of an all-reflective cavity
concept based on low 1st order diffraction efficiency gratings
that was successfully used to construct a high finesse cavity.

The paper is organized as follows: after a short summary of the
basic principles of grating beam splitters and all-reflective
interferometer concepts a theoretical description of a cavity
concept in 2nd order Littrow mount is given.
The design and fabrication of the grating is explained briefly
followed by a comparison of the experimental cavity properties
with theoretical results.
\section{Basic all-reflective interferometer concepts}
Transmissive beam splitters are traditionally used to split and
recombine optical beams in interferometers.
If transmission through optical substrates is unfavorable,
reflection gratings can serve as beam splitters.
\begin{figure}[h]\centerline{\scalebox{.95}{\includegraphics{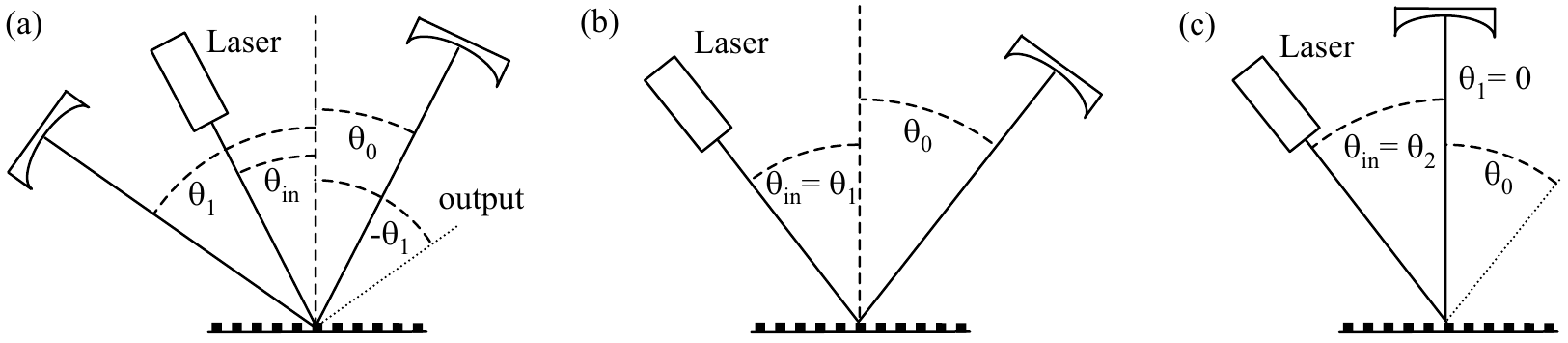}}}
  \caption{(a) grating in non Littrow configuration with
  two existing orders can be used as a beam splitter for a Michelson
  interferometer; (b) linear grating Fabry-Perot interferometer in
  1st order Littrow mount; (c) 2nd order Littrow mount.}
   \label{fig:basic}
\end{figure}
For a laser beam of wavelength $\lambda$ incident onto a grating,
the output angle of the $m$th diffracted order is given by the
grating equation
\begin{equation}
    d(\sin\theta_m + \sin \theta_{\mathrm{in}})= m\lambda,
\end{equation}
where $d$ is the grating period and $\theta_{\mathrm{in}}$ is the
angle of incidence.
The number of existing diffraction orders depends on the choice of
$d,\lambda$ and $\theta_{\mathrm{in}}$.

To obtain an analog to a transmissive beam splitter the parameters
are chosen such that only one additional diffraction order to the
$m=0$ order is present (see Fig. 1(a)).
Michelson and Sagnac interferometers can be formed when the
grating is used in a non Littrow mount and the power of an
incoming beam is split equally into the two orders.

A linear Fabry-Perot interferometer with a coupler in 1st order
Littrow mount
$(\theta_{\mathrm{in}} = \theta_{1})$
is formed when a mirror is placed to retro-reflect the 0th order
to the grating (see Fig. 1(b)).
The finesse of such a cavity is limited by the 1st order
diffraction efficiency of the grating.
If a grating is used in 2nd order Littrow mount and a mirror is
used to retro-reflect the 1st order, the finesse of the resulting
linear Fabry-Perot cavity (see Fig. 1(c)) is only limited by the
reflectance of the grating for normal incidence.
Since high reflectance values are unequally easier to achieve than
high diffraction efficiency values, 2nd order Littrow mounting is
likely to be the more appropriate concept for all-reflective
coupling to high-finesse Fabry-Perot interferometers.
\section{All-reflective Fabry-Perot cavity in 2nd order Littrow mount}
Conventional transmitting beam splitters always couple one input
beam to two output beams.
The input-output relations of a conventional two mirror
Fabry-Perot interferometer follow directly from the phase relation
of the reflected and transmitted beams.
If the length $L$ of the cavity is expressed as a tuning
$\phi=\omega L/c$, where $\omega$ is the angular frequency of the
light and $c$ is the speed of light, the amplitude reflectance
$r_{\mathrm{FP}}$ and transmittance $t_{\mathrm{FP}}$ of a cavity
can be written as
\begin{eqnarray}
r_{\mathrm{FP}}&=& [\rho_0 -\rho_1\exp(2i\phi)]d\,,\\
t_{\mathrm{FP}}&=& -\tau_0\tau_1 \exp(-i\phi) d\,,
\end{eqnarray}
where $\rho_{0,1}$ and $\tau_{0,1}$ denote the reflectance and
transmittance of the two cavity mirrors, respectively, and we have
introduced the resonance factor
$d =[1-\rho_0\rho_1\exp(2i\phi)]^{-1}\,.$

\begin{figure}[h]\centerline{\scalebox{.85}{\includegraphics{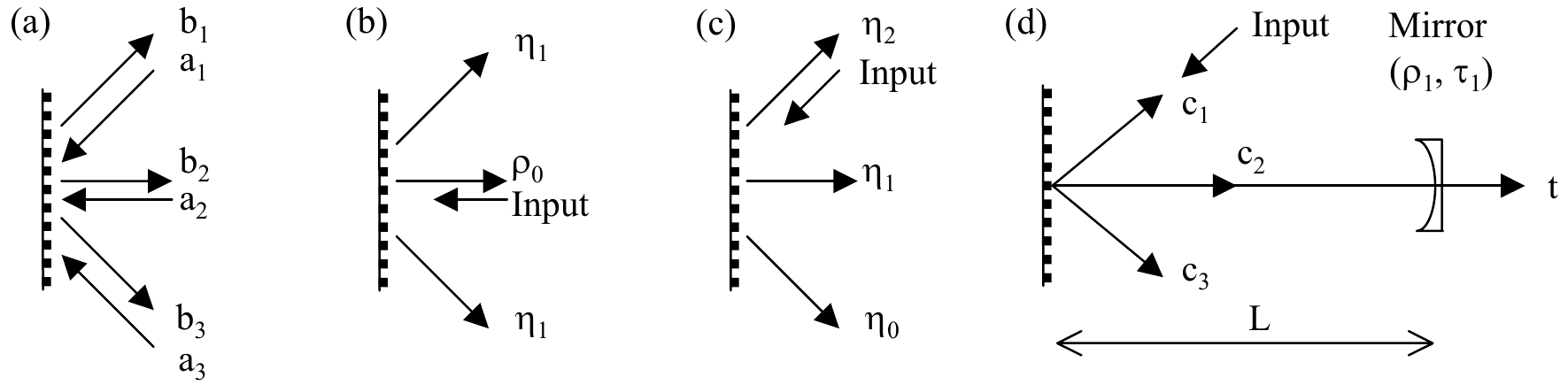}}}
  \caption{3-port reflection grating: (a) labelling of the input
  and output ports; (b) amplitudes of reflection coefficients for
  normal incidence; (c) for 2nd order Littrow incidence; (d) grating
  cavity in 2nd order Littrow mount and the amplitudes of  back
  reflected light $c_1$, intra-cavity field $c_2$, forward reflected
  light $c_3$ and the transmitted light $t$.}
   \label{fig:names}
\end{figure}

In contrast to a transmissive beam splitter the 2nd order Littrow
grating beam splitter couples one input always to three outputs.
For normal incidence it couples to the orders -1, 0, +1 and for
2nd order Littrow incidence
$\theta_{\mathrm{in}}=\arcsin(\lambda/d)$
it couples to the orders 0, 1, 2.
The corresponding amplitude diffraction efficiencies are termed
$\eta_1, \rho_0, \eta_1$ and $\eta_0, \eta_1, \eta_2$,
respectively, as depicted in Fig.~\ref{fig:names}.
The -1st and 1st order for normal incidence have the same
diffraction coefficient for a symmetric grating structure.
For loss less gratings $\rho_0^2+2\eta_1^2=1$ and
$\eta_0^2+\eta_1^2+\eta_2^2=1$ hold.
Coupling to three instead of two output ports results in more
complex phase relations which lead to different cavity properties
compared to a conventional cavity.
The phase relations of the three ports are described by means of a
scattering matrix~\cite{Siegman} formalism in which a complex
valued $3\times 3$ scattering matrix $\mathbf{S}$ represents the 3
port beam splitter.
The 3 input ports are represented by a vector $\mathbf{a}$ with
components $a_i$ that are the complex amplitudes of the incoming
waves at the $i$th port.
The outgoing amplitudes $b_i$ are represented by vector
$\mathbf{b}$.
The input and output ports are coupled via
$\mathbf{b}= \mathbf{S}\times\mathbf{a}\,.$
The grating matrix can be written as~\cite{Bunkowski05}
\begin{equation}
\label{eq:S} {S}_{3p}\!=\!
\left(%
\begin{array}{lcr}
  \eta_2\exp(i\phi_{2}) \!&\! \eta_1\exp(i\phi_{1})   \!&\! \eta_0\exp(i\phi_0)      \!\\
  \eta_1\exp(i\phi_{1}) \!&\! \rho_0\exp(i\phi_0)     \!&\! \eta_1\exp({i\phi_{1}}) \!\\
  \eta_0\exp(i\phi_0)   \!&\! \eta_1\exp(i\phi_{1})   \!&\! \eta_2\exp(i\phi_{2})    \!\\
\end{array}%
\right),
\end{equation}
where $\phi_0, \phi_1, \phi_2$ is the phase shift for 0th, 1st,
2nd order diffraction respectively.
For a loss less grating $\mathbf{S}$ must be unitary and
$|S_{ij}|=|S_{ji}|$ holds for the matrix elements due to
reciprocity of the device.
There is no unique solution for the phases $\phi_i$ since one can
choose different reference planes for the various input and output
ports.
If without loss of generality we assume that specular reflection
is associated with no phase change one gets
\begin{eqnarray}
\phi_0&\!=\!&0\,,\\
\phi_1&\!=\!&-(1/2) \arccos [ (\eta_1^2 - 2\eta_0^2)/(2\rho_0\eta_0)]\,,\\
\phi_2&\!=\!& \arccos[-\eta_1^2/(2\eta_2\eta_0)]. \label{eq:phi2}
\end{eqnarray}
For a given normal incidence reflectivity $\rho_0$ there are
limits for $\eta_0$ and $\eta_2$, namely
\begin{equation}\label{eq:limits}
\eta_{0,\mathrm{_{min}^{max}}}=\eta_{2,\mathrm{_{min}^{max}}}=(1\pm\rho_0)/2.
\end{equation}
It should be noted that these limits are fundamental in the sense
that a reflection grating can only be designed and manufactured
having diffraction efficiencies within these boundaries.

A cavity in 2nd order Littrow mount with an end mirror
reflectivity $\rho_1$, transmittance $\tau_1$ and unity input in
port one, as depicted in Fig.~2 is described by
\begin{equation}
\left(%
\begin{array}{c}
  c_1 \\
  c_2 \\
  c_3 \\
\end{array}%
\right) = {S}_{3p}\times
\left(%
\begin{array}{c}
  1 \\
  \rho_1 c_2\exp({2i\phi}) \\
  0 \\
\end{array}%
\right),
\end{equation}
where $c_1$ is the amplitude of the field reflected back to the
laser, $c_2$ the intra cavity field and $c_3$ the field of the
forward reflected port.
Solving for the amplitudes yields
\begin{eqnarray}\label{eq:amplitudes}
c_1&=& \eta_2\exp(i\phi_2) + \eta_1^2\exp[2i(\phi_1+\phi)]d,\\
c_2&=& \eta_1 \exp(i\phi_1)d,\\\label{eq:amplitude_c3}
c_3&=&\eta_0 + \eta_1^2\exp[2i(\phi_1+\phi)]d\,,\\
t&=&i\tau_1 c_2 \exp(i\phi)
\end{eqnarray}
and $t$ is the amplitude of the light transmitted through the
cavity.

The only grating parameter that determines the finesse of such a
grating cavity is $\rho_0.$
For  given values of $\rho_0$ and $\eta_1$ the intra cavity power
$|c_2|^2$ and the transmitted power $|t|^2$ do not depend on
$\eta_0$ and $\eta_2$.
The power of the two reflection ports  $|c_1|^2$ and $|c_3|^2$
however strongly depend on the values of $\eta_0$ and $\eta_2$.
It is therefore possible to tune the cavity properties of the two
reflecting ports by means of controlling the 0th and 2nd order
diffraction efficiency in the grating production process.
\begin{figure}[h]\centerline{\scalebox{1}{\includegraphics{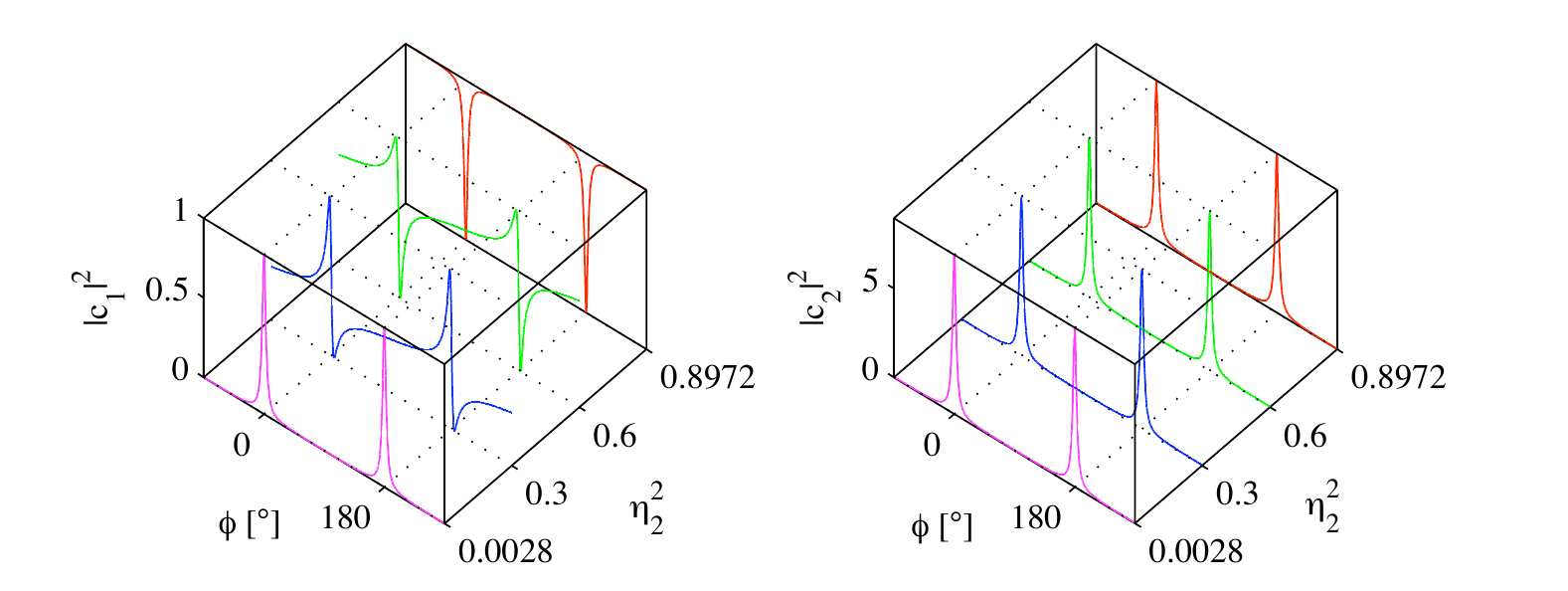}}}
  \caption{left: Power reflectance  $|c_1|^2$ of cavity back reflecting
  port for a grating cavity with end mirror reflectivity $\rho_1=0$ and
  cavity coupling $\eta_1^2=0.1$ for selected values of $\eta_2$; right:
  power inside the cavity $|c_2|^2$.}
  \label{fig:etaminmax}
\end{figure}
Fig.~\ref{fig:etaminmax} illustrates how the power reflectance
$|c_1|^2$ out of the back reflecting port varies as a function of
$\eta_2$ and the tuning $\phi$ of the cavity.
For simplicity a cavity with a perfect end mirror $\rho_1=1$ is
assumed.
The coupling to the cavity is $\eta_1^2=0.1.$
For a coupler with $\eta_2^2= \eta^2_{2,\mathrm{max}}\approx
0.8972$, the cavity does not reflect any light back to the laser
for a tuning of $\phi=0$.
This corresponds to an impedance matched cavity that transmits all
the light on resonance.
For a coupler with $\eta_2^2=\eta^2_{2,\mathrm{min}}\approx
0.0028$, the situation is reversed and all the light is reflected
back to the laser.
For all other values of $\eta_2$ the back-reflected power has
intermediate values and as a significant difference to
conventional cavities: the intensity as a function of
cavity-tuning is no longer symmetric to the $\phi=0$ axis.
\section{Grating design and fabrication}
\begin{figure}[h]\centerline{\scalebox{0.4}{\includegraphics{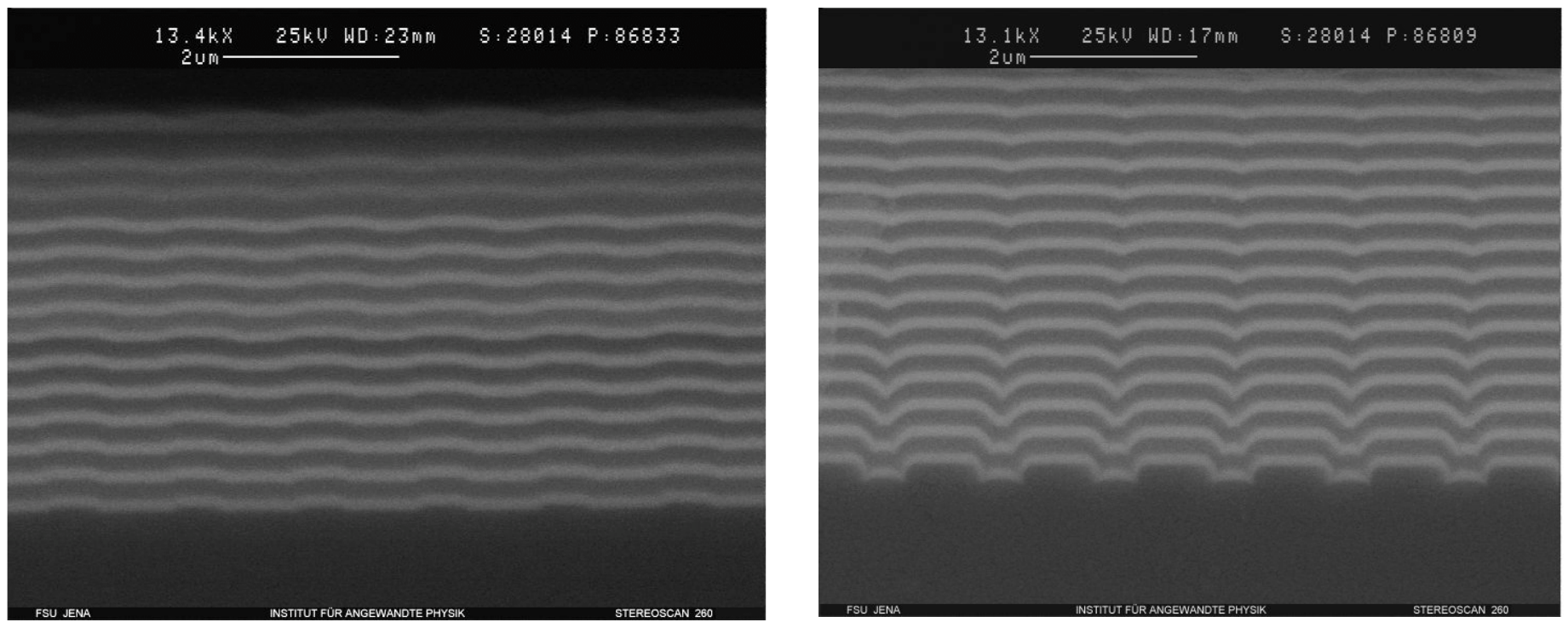}}}
  \caption{Cross sections of overcoated binary gratings (SEM-images)
  with $d=1450\,$nm; left: groove depth of $40-50\,$nm; right: groove
  depth 150\,nm. The rectangular pattern visible at the bottom is
  washed out towards the top of the grating.}
   \label{pic}
\end{figure}

The dielectric grating used as 3-port input coupler unite low
diffraction efficiency and high reflectivity in a single
component.
A common approach to produce dielectric high diffraction
efficiency grating is to etch a periodic structure in the top
layer of a dielectric multilayer stack~\cite{Shore96}.
Here we useed a different approach.
We first etched the grating into a  fused silica substrate and
then overcoated it
such that the dielectric layers effectively form a volume grating
as can be seen in Fig.~\ref{pic}.
A grating period of $d= 1450\,$nm was used corresponding to a 2nd
order Littrow angle $\theta_{\mathrm{in}}=\lambda/d \approx
47.2^{\circ}$ for the Nd:Yag laser wavelength of
$\lambda=1064\,$nm used.
A shallow binary structure with a depth of 40-50\,nm, a ridge
width of 840\,nm was generated by electron beam lithography and
reactive ion beam etching on top of a fused silica substrate.
The applied multilayer stack was composed of 32 alternating layers
of silica (SiO$_2$) and tantalum pentoxide (Ta$_2$O$_5$).
We used a power meter to measure a diffraction efficiency of
$\eta_1^2=0.58\%$ and $\eta_2^2=0.13\%$  for 1064\,nm light with a
polarization plane parallel to the grating grooves and
perpendicular to the plane of incidence ($s$-polarization).
The normal incidence reflectivity $\rho_0^2=98.5\%$ was estimated
via a finesse measurement of a grating cavity consisting of the
grating and an end mirror with known
reflectivity~\cite{Bunkowski}.
\section{Experimental results}
\begin{figure}[h]\centerline{\scalebox{0.85}{\includegraphics{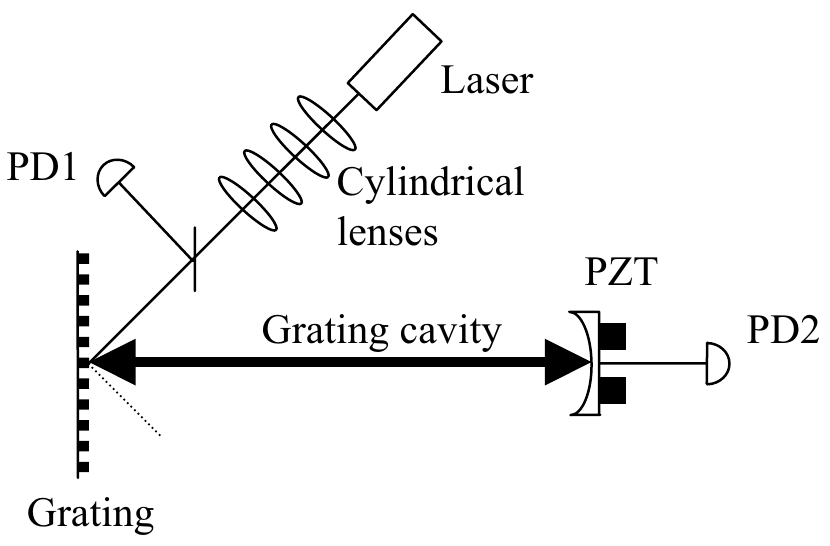}}}
  \caption{Experimental setup for the demonstrated grating Fabry-Perot cavity: PZT, piezoelectric transducer; PD, photo diode.}
   \label{fig:setup}
\end{figure}

Fig.~\ref{fig:setup} shows the experimental setup for the
all-reflective Fabry-Perot cavity.
An end mirror with $\rho_1^2\approx 0.99$ and a radius of
curvature of 1.5\,m mounted on a piezoelectric transducer (PZT) to
allow for cavity length control was placed parallel to the grating
surface at a distance of 43\,cm.
An $s$-polarized beam of 50\,mW from a 1.2\,W, 1064\,nm diode
pumped Nd:YAG laser was used.
Photo detector PD1 is used to monitor the back reflected light
from the cavity and PD2 is used to monitor the light transmitted
through the cavity
To match the eigenmode of the grating cavity the incoming beam
must have an elliptical beam profile which is generated by two
pairs of cylindrical lenses.

\begin{figure}[h]\centerline{\scalebox{1}{\includegraphics{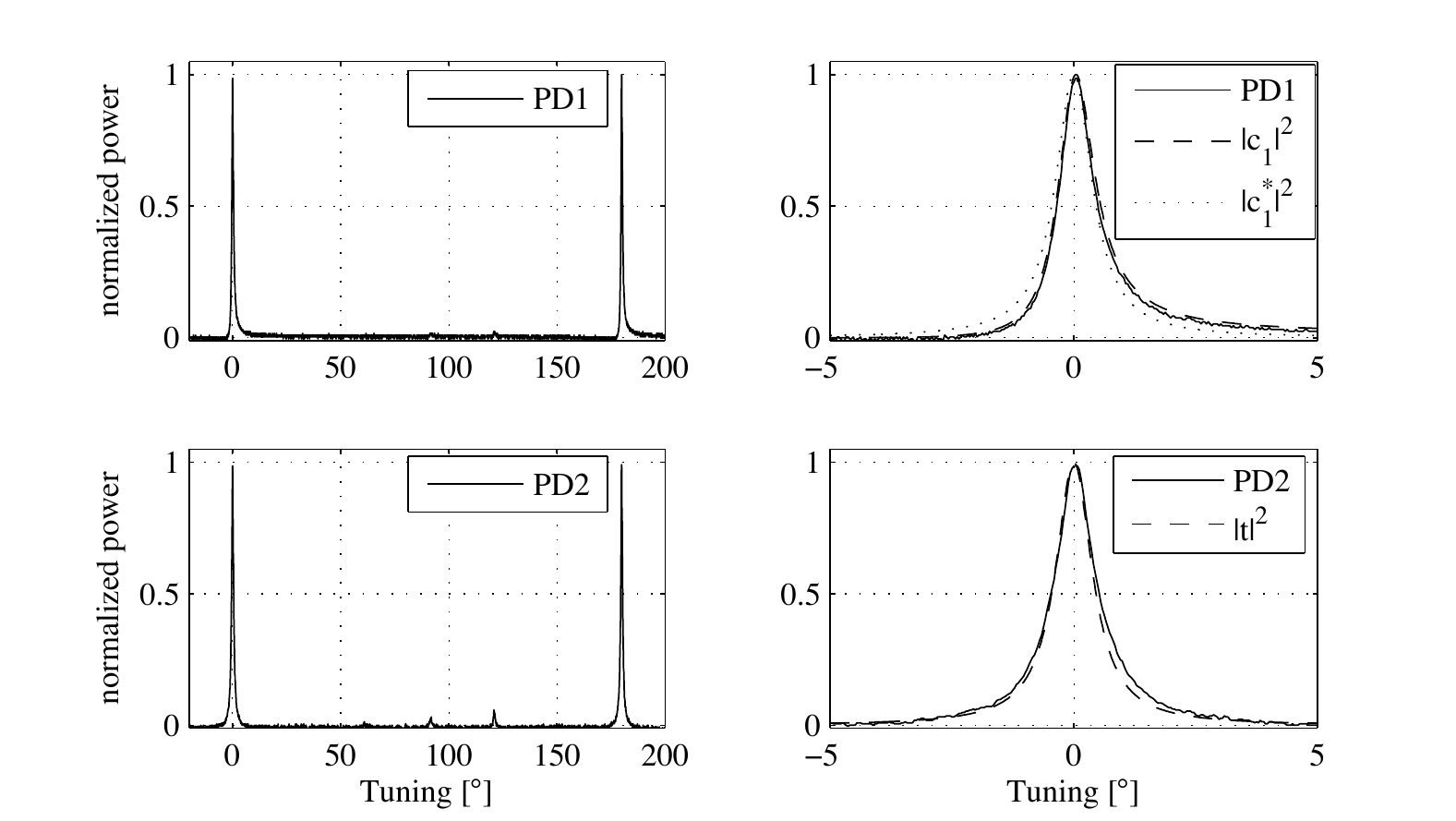}}}
  \caption{Normalized Measured power at PD1 and PD2 and corresponding theoretical curves.
}
   \label{fig:data}
\end{figure}

Fig.~\ref{fig:data} shows measured PD signals normalized to unity
as the cavity length is scanned.
On the left hand side a scan over one free spectral range of the
cavity is shown.
The right hand side is a zoom around one resonance peak.
%
For comparison the normalized theoretical curves for $|c_1|^2$ and
$|t|^2$ are shown.
The lower curve is the well known transmission peak of a Fabry
Perot cavity symmetric to the $\phi=0$ axis.
The upper curve however is not symmetric to zero tuning.
To emphasize the asymmetry we have also plotted
$|c_1^*|^2\equiv|c_1(\phi,\eta_{\mathrm{2,min}})|^2$ which is the
reflected power if $\eta_2$ had its minimal allowed value of
$\eta_{\mathrm{2,min}}$.
Note that the observed asymmetry is not as pronounced as for the
exemplary curves in Fig. \ref{fig:etaminmax} since $\eta_2$ is
relatively close to $\eta_{\mathrm{2,min}}$.
The theoretical curves agree well with the experimental data which
confirms the input-output relations of the grating cavity in 2nd
order Littrow mount.
\section{Conclusion}
The input-output relations of an all-reflective cavity concept
that relies on low diffraction efficiency gratings only have been
derived.
A first test using a shallow dielectric grating  with a groove
depth of 40-50 nm experimentally confirms these relations.
For a complete test we will design and manufacture several
gratings with constant $\eta_1$ but different values for $\eta_2$
and $\eta_0$ thereby tuning the properties of the two reflected
ports.

Additionally we will dedicate more research to the design and
manufacturing of all-reflective cavity couplers for 1st order
Littrow mount and 50/50 beam splitters.
The reduction of the overall optical loss of the gratings due to
transmission and scattering~\cite{Clausnitzer} and a precise
control of the diffraction efficiency of the various diffraction
orders are our main goals since they are two requirements for
using diffractive optics in future generations of gravitational
wave detectors.
Moreover we will investigate new grating interferometer topologies
as well as practical implementation issues.
\section*{Acknowledgements}
The authors would like to thank the Sonderforschungsbereich TR7 of
the DFG.

\end{document}